\documentclass[twocolumn,amsmath,amssymb,showpacs]{revtex4}
\usepackage{bm}
\usepackage{graphicx}
\usepackage{epstopdf}

\begin{document}

\newcommand{\uu}[1]{\underline{#1}}
\newcommand{\pp}[1]{\phantom{#1}}
\newcommand{\be}{\begin{eqnarray}}
\newcommand{\ee}{\end{eqnarray}}
\newcommand{\ve}{\varepsilon}
\newcommand{\vs}{\varsigma}
\newcommand{\Tr}{{\,\rm Tr\,}}
\newcommand{\pol}{\frac{1}{2}}

\title{
Crystallization of space: Space-time fractals from fractal arithmetic}
\author{Diederik Aerts$^1$, Marek Czachor$^{1,2}$, and Maciej Kuna$^{1,2}$}
\affiliation{
$^1$ Centrum Leo Apostel (CLEA),
Vrije Universiteit Brussel, 1050 Brussels, Belgium,\\
$^2$ Wydzia{\l} Fizyki Technicznej i Matematyki Stosowanej,
Politechnika Gda\'nska, 80-233 Gda\'nsk, Poland
}

\begin{abstract}
Fractals such as the Cantor set can be equipped with intrinsic arithmetic operations (addition, subtraction, multiplication, division) that map the fractal into itself. The arithmetic allows one to define calculus and algebra intrinsic to the fractal in question, and one can formulate classical and quantum physics within the fractal set. In particular, fractals in space-time can be generated by means of homogeneous spaces associated with appropriate Lie groups. The construction is illustrated by explicit examples.
\end{abstract}
\maketitle

\section{Introduction}

There are various reasons why fractal sets in space-time are intriguing. For example, quantum gravity \cite{Benedetti,Modesto,COT,QG1,QG2,QG3} or causal trangulation theory \cite{Ambjorn} suggest that space-time itself might posses certain fractal features --- either at small distances, or at early phases of the Universe. At the other extreme are all those astrophysics or cosmological problems where one encounters fractal-like sets embedded in a non-fractal space-time. The typical examples include fractal aspects of galaxies, cosmic voids, or dark matter halos \cite{Piet1,Piet2,Sylos,Balian,Gaite,Bagla,Cordona}.

On the other hand, one can regard a putative space-time fractality as the very origin of quantum physics. Here one should mention Hausdorff 2-dimensionality of Feynman paths \cite{Abbott}, or Ord's derivation of uncertainty and de Broglie relations from 2-dimensional fractal trajectories \cite{Ord}. The scale-relativity project of Nottalle \cite{Nottale84,Nottale93,Nottale} leads to quantum mechanics as a version of mechanics in a non-differentiable, fractal space-time.
A similar philosophy can be found in studies on diffusion on fractals \cite{Barlow}, and their culmination in analysis \cite{Kigami} and differential equations \cite{Strichartz} on non-smooth spaces. It is quite typical to associate fractality of space-time with fractional differential structures \cite{Podlubny,FC1,FC2,FC3}.

In the present paper we follow an alternative approach \cite{MC}. The departure point is to find arithmetic operations that map the fractal in question into itself. Once one knows how to add, multiply, subtract and divide elements of the fractal, one automatically obtains appropriate derivatives, integrals, differential equations, group representations, and thus practically all ingredients needed for classical and quantum physics. Fractal subsets of space-time are then generated by means of homogeneous spaces associated with Lie groups whose parameters satisfy fractal arithmetic. Fractals equipped with arithmetic operations possess intrinsic Lie symmetries that are easy to overlook if one does not have control over the arithmetic.

It is particularly striking that the formalism creates a room for continuous physical processes occurring in sets of zero measure \cite{MC}. For example, quantum harmonic oscillations in the Cantor set are invisible from the point of view of quantum mechanics since quantum states are insensitive to modifications of Schr\"odinger wave functions on sets of zero Lebesgue measure. Still, one can solve the Schr\"odinger equation in the Cantor set and find the energy eigenstates. The resulting energy is physically analogous to dark energy, as it literally `comes out of nowhere' from the point of view of quantum mechanics.

The goal of the present paper is to explicitly analyze examples of fractal sets that go beyond the simple triadic Cantor set discussed in \cite{MC}. We begin with a representation of real numbers where the standard fixed base (binary, triadic...) is replaced by a sequence of probabilities representing different local resolutions of the real line. Having a generalization of the triadic representation we can define an appropriate generalization of the Cantor set equipped,  by construction, with its own intrinsic arithmetic.

We illustrate general considerations with explicit plots of 2-dimensional structures generated by means of rotations in Cantorian plane and Lorentz transformations in 1+1 dimensional Cantorian Minkowski space. The resulting sets possess symmetries inherited from the group that generates the homogeneous spaces, although in order to reveal the symmetries one first has to understand the arithmetic behind them.

The formalism one arrives at is mathematically simple and surprisingly rich, but many interpretational questions remain. The term `crystallization of space' has been inspired by the art of Ludwika Ogorzelec.

\section{Fractal arithmetic and symmetries}

Following the general formalism from \cite{MC} we define
\be
x\oplus y &=& f^{-1}\big(f(x)+f(y)\big),\nonumber\\
x\ominus y &=& f^{-1}\big(f(x)-f(y)\big),\nonumber\\
x\odot y &=& f^{-1}\big(f(x)f(y)\big),\nonumber\\
x\oslash y &=& f^{-1}\big(f(x)/f(y)\big),\nonumber
\ee
where $x,y\in X$, and $f:X\to Y\subset\mathbb{R}$ is a bijection. In later applications we will basically concentrate on an appropriately constructed fractal $X$, but the results are more general. This is an example of a non-Diophantine arithmetic in the sense of \cite{Burgin}.

One verifies the standard properties: (1) associativity $(x\oplus y)\oplus z=x\oplus (y\oplus z)$,
$(x\odot y)\odot z=x\odot (y\odot z)$, (2) commutativity $x\oplus y=y\oplus x$, $x\odot y=y\odot x$, (3) distributivity
$(x\oplus y)\odot z=(x\odot z)\oplus (y\odot z)$. Elements $0,1\in X$ are defined by $0\oplus x=x$, $1\odot x=x$, which implies $f(0)=0$, $f(1)=1$.
One further finds $x\ominus x=0$, $x\oslash x=1$, as expected.
A negative of $x\in X$ is defined as $\ominus x=0\ominus x=f^{-1}\big(-f(x)\big)$, i.e.
$f(\ominus x)=-f(x)$ and $f(\ominus 1)=-f(1)=-1$, or $\ominus 1=f^{-1}(-1)$.
Note that
\be
(\ominus 1)\odot(\ominus 1)
&=&
f^{-1}\Big(f(\ominus 1)^2\Big)=f^{-1}(1)=1.
\ee
In general, one has to be careful to distinguish unit elements occurring at both sides of $f(1)=1$. For example, the rescaled-multiplication approach of Benioff \cite{Benioff,Benioff2} can be regarded as a particular case of the above formalism with $f(x)=px$, $p\neq 0$. Indeed, $x\odot y=(1/p)(pxpy)=pxy$, $x\oplus y=(1/p)(px+py)=x+y$, $x\oslash y=(1/p)(px)/(py)=x/(py)$, but $f(1/p)=1$. Since
$(1/p)\otimes x=(1/p)\big(p(1/p)px\big)=x$ one infers that $1_p=f^{-1}(1)=1/p$ is the unit element of multiplication in Benioff's non-Diophantine arithmetic.

Now, let $F:\mathbb{R}\to \mathbb{R}$ and $F_f:X\to X$ be related by
\be
F_f(x) = f^{-1}\Big(F\big(f(x)\big)\Big).\label{F_f}
\ee
As noted in \cite{MC} the trigonometric functions $\sin_f x=f^{-1}\big(\sin f(x)\big)$, $\cos_f x=f^{-1}\big(\cos f(x)\big)$ satisfy the standard trigonometric formulas, provided `plus' and `times' are represented by $\oplus$ and $\odot$. In consequence
\be
x' &=& x \odot \cos_f \alpha\oplus y \odot \sin_f \alpha,\label{rot1}\\
y' &=& y \odot \cos_f \alpha \ominus x \odot \sin_f \alpha,\label{rot2}
\ee
is a rotation in $X^2$. Fig.~1 shows circles of various radii, defined parametrically by
\be
X\ni \alpha \mapsto (r\odot \cos_f\alpha,r\odot \sin_f\alpha)\in X^2.
\ee
The bijection $f$ employed in Fig.~1 corresponds to the standard triadic Cantor line $X=C$ \cite{MC} and for simplicity is taken in the antisymmetric form $f(-x)=-f(x)$ (let us stress again that in general $-x$ and $\ominus x$ cannot be identified).
The circles are examples of fractal homogeneous spaces, here corresponding to the rotation group in $X^2$. Homogeneous spaces of the (1+1)-dimensional Lorentz group are the hyperbolas,
\be
X\ni \alpha \mapsto (r\odot \cosh_f\alpha,r\odot \sinh_f\alpha)\in X^2,
\ee
depicted in Fig.~2 (with the same $f$ as in Fig.~1, and with hyperbolic functions of the form (\ref{F_f})).

Analogues of arithmetic-based higher dimensional space-time fractals can be found in Ludwika Ogorzelec's installations from her `Crystallization of space' cycle (Fig.~3) \cite{LO1,LO2}.

\begin{figure}
\includegraphics[width=7cm]{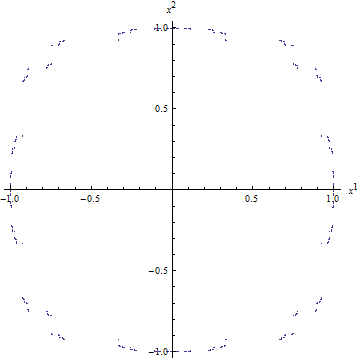}\\
\includegraphics[width=7cm]{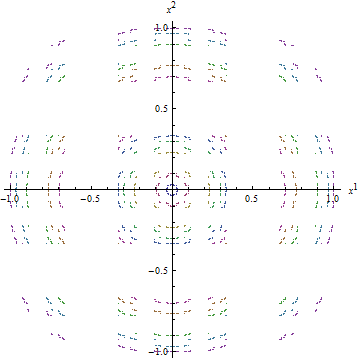}\\
\includegraphics[width=7cm]{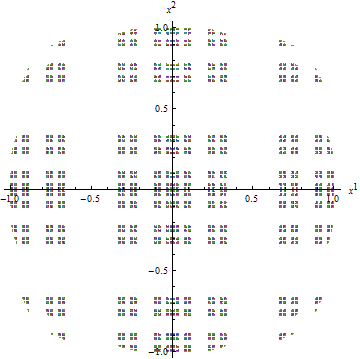}
\caption{A new type of rotational symmetry. Circles in Cantorian plane $C^2$ for various radii. From top to bottom: 1, 10, and 50 circles.}
\end{figure}

\begin{figure}
\includegraphics[width=7cm]{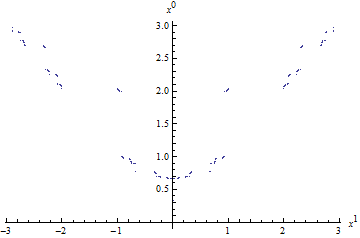}\\
\includegraphics[width=7cm]{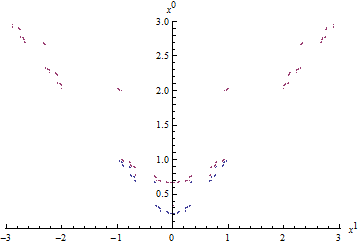}\\
\includegraphics[width=7cm]{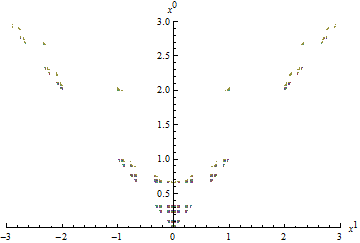}
\caption{Hyperbolic symmetry. Proper-time hyperbolas in (1+1)-dimensional Cantorian Minkowski space-time $C^2$. From top to bottom: 1, 2, and 20 hyperbolas.}
\end{figure}
\begin{figure}
\includegraphics[width=4cm]{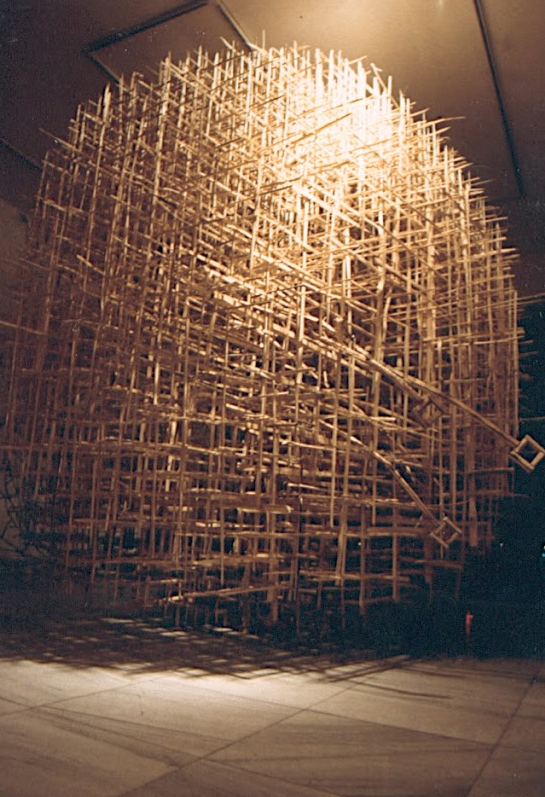}
\caption{An Ogorzelec set. Higher dimensional space-time structures generated by fractal arithmetic are analogous to L.~Ogorzelec installations from her `Crystallization of space' cycle. Mathematical Ogorzelec sets can be constructed by foliations consisting of homogeneous spaces of a symmetry Lie group with parameters subject to fractal arithmetic.}
\end{figure}

\section{Fractal derivatives and integrals}

A derivative of a function $A:X\to X$ is defined by
\be
\frac{d_f A(x)}{d_fx}
&=&
\lim_{h\to 0}\Big(A(x\oplus h)\ominus A(x)\Big)\oslash h.\label{der}
\ee
It satisfies
\be
\frac{d_f A(x)\odot B(x)}{d_fx}
&=&
\frac{d_f A(x)}{d_fx}\odot B(x)
\oplus
A(x)\odot \frac{d_fB(x)}{d_fx},\nonumber\\
\frac{d_f A(x)\oplus B(x)}{d_fx}
&=&
\frac{d_f A(x)}{d_fx}
\oplus
\frac{d_f B(x)}{d_fx},
\nonumber\\
\frac{d_f A[B(x)]}{d_fx}
&=&
\frac{d_f A[B(x)]}{d_f B(x)}
\odot
\frac{d_f B(x)}{d_fx}.\nonumber
\ee
Employing (\ref{der}) and the fact that $f(0)=0$ one finds for functions of the form (\ref{F_f})
\be
\frac{d_f F_f(x)}{d_fx}
&=&
f^{-1}\Big(F'\big(f(x)\big)\Big),\label{derf}
\ee
where $F'(y)=dF/dy$ is the usual derivative in $Y$, defined in terms of $+$, $-$, $\cdot$, and $/$. Note that no derivatives of $f$ and $f^{-1}$ occur in (\ref{derf}). In particular, $d_f \exp_f x/d_fx=\exp_f x$, $d_f \cos_f x/d_fx=\ominus\sin_f x$, $d_f \sin_f x/d_fx=\cos_f x$, and so on.

There is no relation between $d_f/d_fx$ and fractional derivatives. In fact, one could formulate non-Diophantine analogs of fractional derivatives and integrals, if needed. In order to do this one simply has to know how to integrate in a way guaranteeing the standard laws of the calculus.

An integral is defined so that the fundamental laws,
\be
\int_a^b \frac{d_f A(x)}{d_fx}\odot d_fx = A(b)\ominus A(a),\nonumber
\ee
and
\be
\frac{d_f}{d_f x}\int_a^x A(x')\odot d_fx' = A(x),
\ee
hold true. For $F_f(x)$ given by (\ref{F_f}) the explicit form of the integral reads
\be
\int_a^b F_f(x)\odot d_fx = f^{-1}\left(\int_{f(a)}^{f(b)}F(y)dy\right),\label{int}
\ee
where $\int F(y)dy$ is the standard (say, Lebesgue) integral in $\mathbb{R}$.

In this way one arrives at the calculus which is as simple as the one one knows from undergraduate education, and yet one can formulate and solve problems formulated entirely within fractal sets. In practice, the only problem to solve for a given fractal is to find the bijection $f$.

\section{Fractal space-time trajectories}

Not only is Fig.~1 an illustration of fractal homogeneous spaces, but it simultaneously shows phase-space trajectories of a classical nonrelativistic harmonic oscillator in (1+1)-dimensional Cantorian space time \cite{MC}. Fig.~2 shows proper-time hyperbolas defined by
\be
g_{\mu\nu}\odot x^\mu\odot x^\nu
&=&
x^0\odot x^0\ominus  x^1\odot x^1
=
s^{\odot 2}.
\ee
Note that in (1+1)-dimensional Minkowski space one finds
\be
x_0 &=&x^0=g_{00}\odot x^0=1\odot x^0,
\ee
so $g_{00}=1$ is the neutral element of multiplication.
Now, on one hand,
\be
x_1 &=&g_{11}\odot x^1=
f^{-1}\Big(f(g_{11})f(x^1)\Big).
\ee
Putting it differently we find
\be
x_1 &=&\ominus x^1=0\ominus x^1
=
f^{-1}\Big(f(0)-f(x^1)\Big).
\ee
Accordingly,
$f(g_{11})f(x^1)=-f(x^1)$,  $g_{11}=f^{-1}(-1)$.  In general,
\be
x_1 &=&
f^{-1}\Big(-f(x^1)\Big)
\neq -x^1.
\ee
The Lorentz transformations, defined by
\be
x'{^0} &=& x^0 \odot \cosh_f \alpha\ominus x^1 \odot \sinh_f \alpha,\label{Lor1}\\
x'{^1} &=& x^1 \odot \cosh_f \alpha\ominus x^0 \odot \sinh_f \alpha,\label{Lor2}
\ee
satisfy
\be
g_{\mu\nu}\odot x'{^\mu}\odot x'{^\nu}
&=&
g_{\mu\nu}\odot x^\mu\odot x^\nu.
\ee
The characteristic Cantor-like structure visible at the lowest plot at Fig.~2 could be equivalently generated by plotting a bunch of `straight'
world-half-lines, as shown in Fig.~4,
\be
x^\mu(s) &=& u^\mu \odot s,\quad 0\leq s.\label{x(s)}
\ee
The uppermost plot involves only three world-half-lines, two null and one timelike. The null lines look `ordinary', i.e. comply with the intuitive picture of a straight line. The timelike world-line is also `straight' in the sense of formula (\ref{x(s)}), and for inhabitants of Cantorian Minkowski space would appear as `ordinarily straight' as generators of the light cone.

By definition of the derivative we find
\be
\frac{d_fx^\mu(s)}{d_fs}= \lim_{h\to 0}
\Big(u^\mu \odot (s\oplus h)\ominus u^\mu\odot s\Big)\oslash h=u^\mu
\ee
and thus the straight line (\ref{x(s)}) is a space-time trajectory in the usual sense, with four-velocity $u^\mu$. Such a simple family of world-lines is enough to formulate a fractal analogue of the twin paradox.
\begin{figure}
\includegraphics[width=7cm]{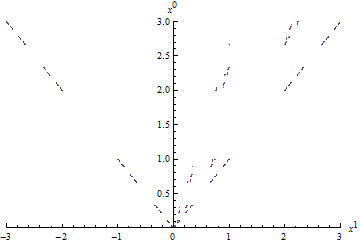}\\
\includegraphics[width=7cm]{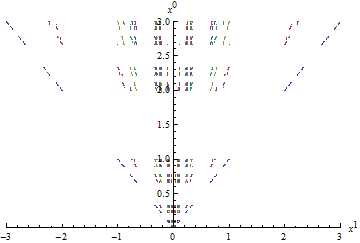}\\
\includegraphics[width=7cm]{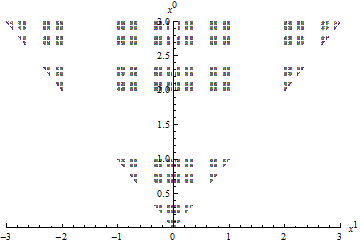}
\caption{World half-lines in (1+1)-dimensional Cantorian Minkowski space-time $C^2$. From top to bottom: Light cone plus 1, 22, and 400 timelike worldlines. The same $f$ as in Fig.~1 and Fig.~2.}
\end{figure}

\section{Fractal Minkowski coordinates}

Lorentz transformations (\ref{Lor1})--(\ref{Lor2}) define coordinate axes as the world-lines
\be
x^0 =\beta\odot x^1 ,\label{Lor10}
\ee
(Lorentz-transformed simultaneity hyperplane of the event $x_\mu=x'_\mu=0$) and
\be
\beta\odot x^0 =x^1 ,\label{Lor20}
\ee
(Lorentz-transformed time axis), where $\beta=\tanh_f \alpha=\sinh_f \alpha\oslash \cosh_f \alpha$.

The left part of Fig.~5 shows two coordinate systems in fractal Minkowski space $C^2$ (see Sec.~VIII). Coordinate axes correspond to $\beta=0$ (vertical and horizontal axes) and $\beta=f^{-1}(1/2)$ (diagonal broken lines), together with the light cone defined by $\beta=1$ and $\beta=\ominus 1$. The right part shows the result of applying $f$ to $C^2$. The bijection $f$ is taken in the irregular `multi-resolution' form, described in detail in Sec.~XI.
\begin{figure}
\includegraphics[width=4cm]{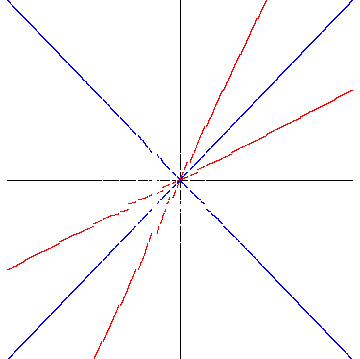}
\includegraphics[width=4cm]{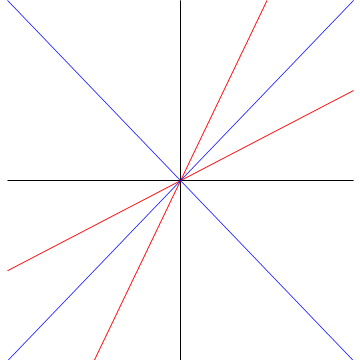}
\caption{Left: fractal light cone and two fractal coordinate systems for the multi-resolution Cantor-line-Minkowski 1+1 dimensional space $C^2$, defined by means of $f$ from Sec.~XI. Right: the same lines but mapped by $f$ into the usual Minkowski space.}
\end{figure}
\section{Twin paradox}

Consider two world lines. The `travelling twin' corresponds to
\be
x^\mu(s)
=
\left\{
\begin{array}{cl}
u^\mu\odot s &\textrm{for $0\leq s < s_1$}\\
u^\mu\odot s_1\oplus v^\mu\odot (s\ominus s_1) &\textrm{for $s_1\leq s\leq s_2$}
\end{array}
\right.
.\label{x(s)1}
\ee
The twin `at rest' is described by
\be
\tilde x^\mu(s)
&=&
\Big(u^\mu\odot s_1\oplus v^\mu\odot (s_2\ominus s_1)\Big)\odot s\oslash s_2,\label{x(s)2}
\ee
for $0\leq s\leq s_2$. Here $x^\mu$, $\tilde x^\mu$, $u$, and $v^\mu$ are position and 4-velocity world-vectors, respectively, with $u_\mu\odot u^\mu=v_\mu\odot v^\mu=1$. Since
$x(0)=\tilde x(0)$ and $x(s_2)=\tilde x(s_2)$ the two trajectories can be used to derive the paradox. The Cantorian Minkowski-space length of $s\mapsto x(s)$ is $S=s_1\oplus (s_2\ominus s_1)=s_2$ whereas the one of $\tilde x(s)$ satisfies $\tilde S^{\odot 2}= g_{\mu\nu}\odot \tilde x^\mu(s_2)\odot \tilde x^\nu(s_2)$. Assume for simplicity that $S=s_2=s_1\oplus s_1=f^{-1}(2)\odot s_1$:
\be
\tilde x^\mu(s)
&=&
(u^\mu\oplus v^\mu)\odot s_1\odot s\oslash s_2,\nonumber\\
\tilde S^{\odot 2}
&=&
(u_\mu\oplus v_\mu)\odot (u^\mu\oplus v^\mu)\odot s_1^{\odot 2}
\nonumber\\
&=&
S^{\odot 2}\odot (1\oplus u_\mu\odot v^\mu)\oslash f^{-1}(2).\label{SS}
\ee
In order to cross-check (\ref{SS}) take the trivial case $f(x)=x$, $(u^0,u^1)=(1,\beta)/\sqrt{1-\beta^2}$,
$(v^0,v^1)=(1,-\beta)/\sqrt{1-\beta^2}$. Then
\be
(1\oplus u_\mu\odot v^\mu)\oslash f^{-1}(2)
&=&\frac{1}{1-\beta^2}=u_0^2\label{u_0^2}
\ee
i.e. $S=\tilde S\sqrt{1-\beta^2}=\tilde S/u^0$, as expected. For a general $f$ let us first note that the normalization
\be
1 &=&
u_\mu\odot u^\mu
\nonumber\\
&=&
u^0\odot u^0\ominus u^1\odot u^1
\nonumber\\
&=&
f^{-1}\Big(f(u^0)^2-f(u^1)^2\Big)
\ee
together with $f(1)=1$ implies $f(u^0)^2-f(u^1)^2=1$. If $v^0=u^0$ and $v^1=\ominus u^1$ then
\be
(1\oplus u_\mu\odot v^\mu)\oslash f^{-1}(2)=f^{-1}\left(f(u^0)^2\right)=u_0^{\odot 2}.\label{fu_0^2}
\ee
(\ref{fu_0^2}) is exactly analogous to (\ref{u_0^2}), so finally we get the simple formula for the time delay which is valid in any $f$-arithmetic Minkowski space,
\be
S=\tilde S\oslash u_0.\label{S odot u}
\ee
Since
\be
1\oslash u_0 =
f^{-1}\left(\sqrt{1-f(\beta)^2}\right),\quad \beta=u^1\oslash u^0,\label{S odot u'}
\ee
we can alternatively write
\be
f(S)=f(\tilde S)\sqrt{1-f(\beta)^2}.\label{S odot u"}
\ee
Comparing $S$ and $\tilde S$ in a space-time neighborhood of a given $x=x(0)=\tilde x(0)$, we can {\it in principle\/} directly probe the form of $f$ (Fig.~6).

The problem is which of the two formulas, (\ref{S odot u}) or (\ref{S odot u"}), should be employed in comparison of experimental data with the theory? Which of the two velocity parameters, $0\leq \beta\leq 1$ or $0\leq f(\beta)\leq 1$, is the one employed in the experiment if we assume that the observers live in the fractal space-time?
Moreover, which $f$ should be employed? The bijection $f$ is non-unique, this is the essence of {\it relativity of arithmetic\/} discussed in \cite{MC}.

It seems at this stage of the formalism we are just lacking appropriate physical intuitions. We do not have yet a `theory of measurement'. Several options are logically possible, so it is best to begin with less trivial examples.

\begin{figure}
\includegraphics[width=7cm]{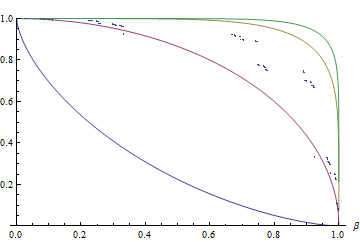}
\caption{The time-delay factor $1\oslash u_0$ of Eq.~(\ref{S odot u'}), for various choices of $f$. The continuous curves correspond to $f(x)=x^q$, with $q=1/3$ (leftmost), $q=1$, $q=3$, and $q=5$ (rightmost). The discrete points correspond to $f$ of the standard triadic Cantor set (as defined in \cite{MC}).}
\end{figure}

\section{Multi-resolution representation of real numbers}

Although Cantor-type sets are homeomorphic to the idealized fully symmetric triadic Cantor set, it is clear that fractal-like sets one encounters in real life are highly non-symmetric and non-regular. Their effective dimensions vary with resolution and are position dependent. The mathematical notion that seems close to natural fractals is associated with the concept of a multifractal. However, in order to apply the idea of fractal arithmetic to a multifractal one needs a bijection $f$, and it is by no means evident that such an $f$ always exists.

So, we propose to reverse the problem. Namely, can we describe a class of fractals that, on one hand, have the irregularities typical of multifractals, but on the other hand are equipped with $f$? The answer is in the affirmative and is related to the concept of a multi-resolution representation of real numbers.

To begin with, let us make the trivial remark that
geometry of physical space-time involves objects that have `dimension of length' ($x$ or $x_0=ct$ are expressed in meters, inches, parsecs, Planck lengths...). In pure mathematics the element $1\in \mathbb{R}$  is just the neutral element of multiplication in the real `line' and, obviously, does not have a `physical unit'. The construction given in \cite{MC} is conceptually in-between these two, `physical' and `mathematical', perspectives. We are interested in physical-space fractals constructed by means of a map $f$ satisfying $f(1)=1$, where the 1s are understood as neutral elements of multiplication. Following the suggestion from \cite{MC} we will treat the physical space as an object which is dimensionless, and this can be obtained only for the price of introducing a fundamental unit of length, $\ell$ say.
\begin{figure}
\includegraphics[width=8cm]{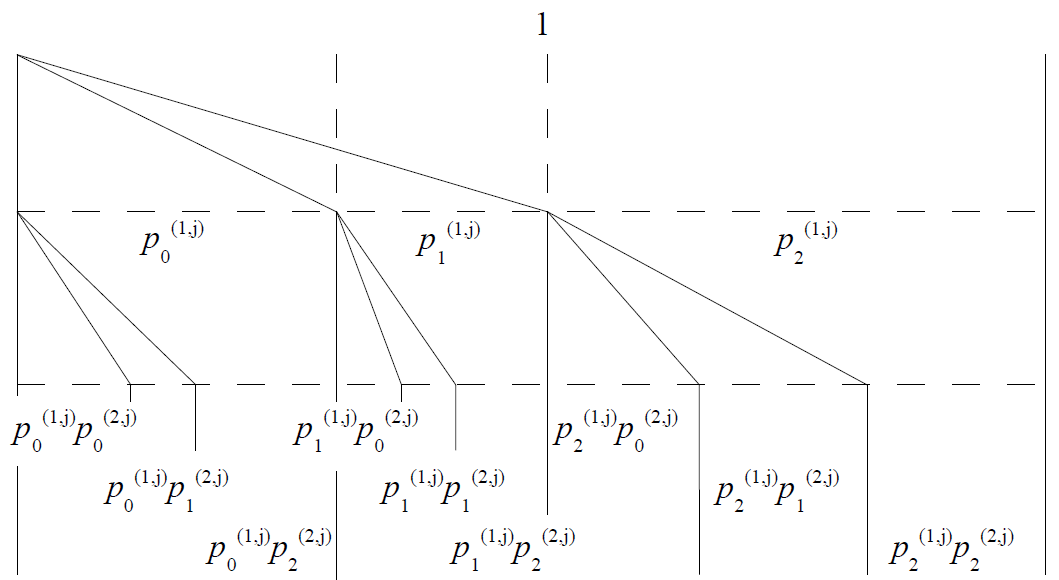}
\caption{A $j$-th interval and its splitting into intervals whose proportions differ from resolution to resolution (indexed by $n$ in $(n,j)$).}
\end{figure}
\begin{figure}
\includegraphics[width=7cm]{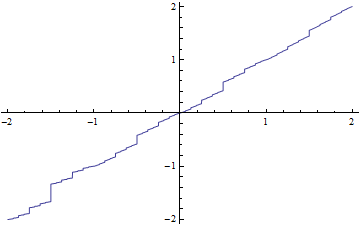}\\
\includegraphics[width=7cm]{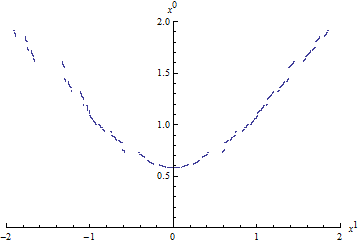}\\
\includegraphics[width=7cm]{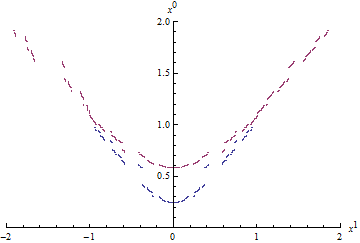}\\
\includegraphics[width=7cm]{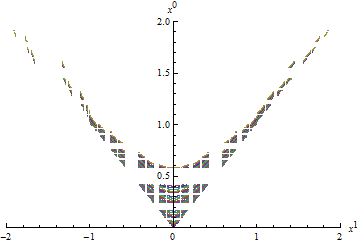}
\caption{Future cone generated by fractal-arithmetic proper-time SO(1,1) homogeneous spaces. From top to bottom: $f^{-1}$, and the corresponding 1, 2, and 20 hyperbolas. Note the lack of exact reflection symmetry, typical of multi-resolution Cantor sets.}
\end{figure}
\begin{figure}
\includegraphics[width=7cm]{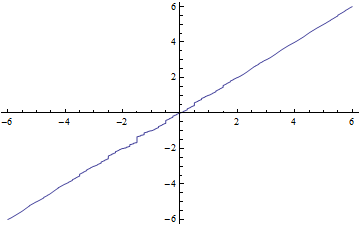}\\
\includegraphics[width=7cm]{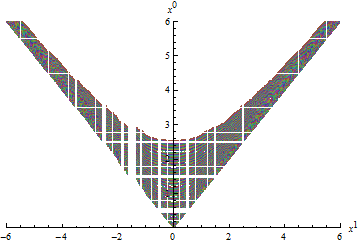}
\caption{The same homogeneous spaces as in Fig.~8 but seen from a larger perspective:  100 proper-time hyperboles. Parity non-invariance is less pronounced at larger length-scales.}
\end{figure}

With this observation in mind let us split a one-dimensional physical `position-space line' $X$ into a countable union of disjoint intervals of length $\ell$. In order to model it mathematically we identify $X/\ell=\mathbb{R}=\cup_{j\in \mathbb{Z}}[j,j+1)$.

The diagram shown in Fig.~7 shows a $j$th interval $[j,j+1)$.
The interval is split into three (right-open) segments of length $ p_k^{(1,j)}$, $k=0,1,2$.
Each of the three segments is yet further split into three right-open intervals whose mutual proportions are determined by $p_k^{(2,j)}$, $k=0,1,2$, and so on.
For each $(n,j)\in \mathbb{N}\times \mathbb{Z}$ we assume $\sum_{k=0}^2p_k^{(n,j)}=1$, $0\leq p_k^{(n,j)}\leq 1$. Denoting
$r_0^{(n,j)}=0$, $r_1^{(n,j)}=p_0^{(n,j)}$, $r_2^{(n,j)}=p_0^{(n,j)}+p_1^{(n,j)}$, and
$s^{j}_{a_1\dots a_n}=p_{a_1}^{(1,j)}\dots p_{a_{n-1}}^{(n-1,j)}r_{a_n}^{(n,j)}$, one can associate each node of the diagram with the real number
\be
x
=
x_{j.a_1\dots a_n}
=
j
+
s^{j}_{a_1}
+
\dots
+s^{j}_{a_1\dots a_n}\in [j,j+1)
,\label{1}
\ee
where $a_k=0,1,2$, $j\in \mathbb{Z}$. There exist two extreme cases of (\ref{1}). Firstly, if all $p_{a_k}^{(n,j)}=1/3$ we obtain a representation where the non-integer part $x-\lfloor x\rfloor$ has the standard ternary form. At the other extreme is the case where the proportions $p_k^{(n,j)}$ are completely unrelated to one another for different choices of $(n,j)$. Of particular interest is, as we shall see later, the intermediate case where $\lim_{n\to\infty} p_k^{(n,j)}=p_k$ exists and is independent of $j$.
We show in the Appendix that
\be
x_{j.a_1\dots a_n1}=x_{j.a_1\dots a_n1(0)}=x_{j.a_1\dots a_n0(2)}.\label{1=0(2)}
\ee
If $x=j\in\mathbb{Z}$, then $x=x_{j.(0)}=x_{j-1.(2)}\in [j,j+1)$. In consequence, there exist numbers that have exactly two different representations (\ref{1}). The set of such numbers is countable.

\section{Multi-resolution Cantor line}

Here we generalize the construction of the Cantor line given in \cite{MC} (for $p_{a_k}^{(n,j)}=1/3$). Our goal is to have a fractal $C\subset \mathbb{R}$ and a bijection $f:C\to \mathbb{R}$ which will be used in definition of fractal arithmetic, an essential ingredient of fractal derivatives and integrals. Fractal $C$ so constructed is, to some extent, reminiscent of a multifractal but, as opposed to standard multifractals, is equipped with the natural bijection $f$. This is why we speak of multi-resolution fractals, distinguishing them from multifractals where arithmetic operations and derivatives are difficult to introduce.

To begin with, consider a real number $y=j+\epsilon\in[j,j+1)$. The $\epsilon\in[0,1)$ has at most two different binary representations,
\be
\epsilon=(0.b_1\dots b_n\dots)_2=(0.b'_1\dots b'_n\dots)_2
\ee
which defines two sequences of bits.
The two sequences allow us to define two numbers of the form (\ref{1}):
\be
x=x_{j.2b_1\dots 2b_n\dots}\textrm{ and }
x'=x_{j.2b'_1\dots 2b'_n\dots},\label{22}
\ee
both belonging to $[j,j+1)$.
Note that out of the three possible digits $a_k=0,1,2$, occurring in (\ref{1}), formula (\ref{22}) involves only two of them: $2b_k=0,2$, $2b'_k=0,2$.
The absence of ternary 1 is typical of Cantor-like sets.

The injective map
$g:\mathbb{R}\to \mathbb{R}$ is defined by $g(y)=\min\{x,x'\}$. The image $C=g(\mathbb{R})$ will be termed the multi-resolution Cantor line. The inverse map
$f:C\to\mathbb{R}$, $f=g^{-1}$, defines the bijection we need in order to construct arithmetic in $C$. Let us check that $f(0)=0$, $f(1)=1$. 0 occurring in the argument of $f(0)$ corresponds to $x_{0.(0)}=x_{-1.(2)}\in C$. By definition $f(x_{0.(0)})=0+0.(0)_2=0$, $f(x_{-1.(2)})=-1+0.(1)_2=0$.
1 in the argument of $f(1)$ corresponds to $x_{1.(0)}=x_{0.(2)}\in C$. Again, by definition $f(x_{1.(0)})=1+0.(0)_2=1$, $f(x_{0.(2)})=0+0.(1)_2=1$.

One similarly shows that $f(j)=f(x_{j.(0)})=j+0.(0)_2=f(x_{j-1.(2)})=j-1+0.(1)_2=j\in \mathbb{Z}$. Thus, for integer $x$ one finds $f(x)=x=f^{-1}(x)$. In particular, $f(-1)=-1$. This is the peculiarity of this concrete $f$, but the general formalism of `relativity of arithmetic' from \cite{MC} requires only that $f(0)=0$ and $f(1)=1$.

\section{Relation to multifractals}

Let us put what we do in the context of multifractals, concentrating only on multifractals of a Cantor type. Assume, first of all, that
$p_a^{(n,j)}=p_a$ for any $(n,j)$, but not all $p_a$ are equal. At resolution $n$ one deals with segments of length $p_0^{n-m}p_2^m$, $m=0,\dots,n$, and each interval $[j,j+1)$ contains $n!/[m!(n-m)!]$ segments of a $m$th type. The overall length of all the segments of the $m$th type is
$p_0^{n-m}p_2^m n!/[m!(n-m)!]$ and the sum over all $m$ is $(1-p_1)^n$. So, if $p_1>0$ then $\lim_{n\to\infty}(1-p_1)^n=0$. Removing in each step a nonzero proportion $p_1$ of $[j,j+1)$ we get in the limit a set of Lebesgue measure zero.

The Hausdorff dimension $D$ is defined by
\be
\sum_{m=0}^n\frac{n!}{m!(n-m)!}\left(p_0^{n-m}p_2^m\right)^D
=(p_0^D+p_2^D)^n=1.
\ee
Hence, $p_0^D+p_2^D=1$ and $D$ coincides with $\lim_{n\to\infty}D^{(n,j)}$ discussed in the next section.

In order to introduce the multifractal formalism \cite{HP1,HP2} one additionally assumes that there exists some random process with probabilities $P_0$, $P_2$, $P_0+P_2=1$ such that the algorithm of generating the fractal may be regarded as a kind of random walk. One introduces a parameter $q$ and a function $\tau(q)$, and demands that
\be
P_0^qp_0^{-\tau(q)}+P_2^qp_2^{-\tau(q)}=1.
\ee
For $q=0$ and $\tau(0)=-D$ one finds that $-\tau(0)$ is the Hausdorff dimension. The so-called generalized dimensions are defined by
$D(q)=\tau(q)/(q-1)$.

Our multi-resolution approach to Cantor-like sets does not naturally lead to any stochastic process of a multifractal type. Moreover, the essential ingredient of the construction from \cite{MC} is the bijection $f$ that leads to arithmetic operations, but there is no natural definition for such a $f$ in the multifractal formalism.

\section{Dimensions of $C$}

With each node $x_{j.a_1\dots a_n}$ from Fig.~7 one can associate the length $l_{j.a_1\dots a_n}$ of the interval extending to the right till its nearest sibling,
\be
l_{j.a_1\dots a_n}
&=&
p_{a_1}^{(1,j)}\dots p_{a_n}^{(n,j)},
\ee
satisfying
\be
\sum_{a_1=0}^2\dots \sum_{a_n=0}^2l_{j.a_1\dots a_n}
=
\prod_{k=1}^n\sum_{a=0}^2p_{a}^{(k,j)}
=1.\label{sum}
\ee
In Cantor-like sets the indices $a=1$ would be missing in sums (\ref{sum}), but one can find numbers $D^{(k,j)}$ such that
\be
\prod_{k=1}^n\sum_{a\neq 1}\left(p_{a}^{(k,j)}\right)^{D^{(k,j)}}
=1.\label{sum'}
\ee
Putting $n=1$ in (\ref{sum'}) we get
\be
\sum_{a\neq 1}\left(p_{a}^{(1,j)}\right)^{D^{(1,j)}}
=1,
\ee
which implies by induction that (\ref{sum'}) is equivalent to
\be
\left(p_{0}^{(k,j)}\right)^{D^{(k,j)}}+\left(p_{2}^{(k,j)}\right)^{D^{(k,j)}}
=1
\ee
which has a unique solution $D^{(k,j)}$ for any $(k,j)$ (the proof is standard; cf. the analysis of similarity dimension in \cite{Edgar}).

Alternatively, one can consider $D_{(n,j)}$ defined by
\be
1 &=&\sum_{a_1\neq 1}\dots \sum_{a_n\neq 1}\left(l_{j.a_1\dots a_n}\right)^{D_{(n,j)}}.
\label{sum'''}
\ee
Eq.~(\ref{sum'''}) possesses a unique solution $D_{(n,j)}$ which, however, in general differs from $D^{(n,j)}$. The limiting case $\lim_{n\to\infty}D_{(n,j)}$ equals the Hausdorff dimension of the $j$th interval.

Dimensions $D^{(n,j)}$ and $D_{(n,j)}$ are the two effective similarity dimensions that can be associated with resolution $n$ in the $j$th segment of $C$. Note that  $D^{(n,j)}=1=D_{(n,j)}$ if and only if $p_1^{(n,j)}=0$, independently of the choice of $p_0^{(n,j)}$ and $p_2^{(n,j)}$. In infinite resolution the dimension $D^{(n,j)}$ is well defined if
$\lim_{n\to\infty}p_a^{(n,j)}$ exists. If the limit does not exist then $D^{(n,j)}$ fluctuates at large resolutions.

Now, let us parametrize the probabilities in a Gibbsian way. Its simplest form reads
\be
p_a^{(n,j)}(T) &=& \frac{e^{-E_a^{(n,j)}/(k T)}}{\sum_{b=0}^2e^{-E_b^{(n,j)}/(k T)}}.
\ee
Assuming $E_0^{(n,j)}<E_1^{(n,j)}<E_2^{(n,j)}$ one finds $p_a^{(n,j)}(\infty)=1/3$ for $a=0,1,2$, and $p_0^{(n,j)}(0)=1$ (hence $p_1^{(n,j)}(0)=0$). The corresponding dimensions are $T$-dependent: $D^{(n,j)}(0)=D_{(n,j)}(0)=1$, $D^{(n,j)}(\infty)=D_{(n,j)}(\infty)=\log_32$.
The change of dimensionality with $T$ can be also expressed in the escort-probability form \cite{Beck,Naudts},
\be
p_a^{(n,j)}(T') &=& \frac{p_a^{(n,j)}(T)^q}{\sum_{b=0}^2 p_b^{(n,j)}(T)^q},\quad q=T/T'.
\ee

In our formalism the space itself, modeled by our $C$, may have properties analogous to matter. One can speak of macro- (small $n$), meso- (intermediate $n$) and micro-structure ($n\gg 1$) of space. Degree of granularity of space is measured in terms of fractal dimensions, but one has to bear in mind that the Hausdorff dimension of a Cartesian product of sets is greater or equal to the sum of Hausdorff dimensions of the sets themselves \cite{Falconer}. Multi-resolution space-time in general does not possess a well defined scaling symmetry and thus it may be difficult to compute its (local) dimensions, since a simple sum may not give the correct result.

The change of dimensionality can be also analyzed in terms of critical phenomena \cite{Crit}. Spontaneous generation of a crystalline ground state in a higher derivative theory \cite{Ghosh} provides a concrete example of such a process. Another related example is provided by the studies of granularity of space in the formalism of path integrals \cite{JF1,JF2}. Path integrals as well as the techniques of signal analysis can be naturally reformulated in the non-Diophantine arithmetic by means of the representation of complex numbers and integration introduced in \cite{MC}. This includes the `momentum' or Fourier representation on fractals equipped with fractal arithmetic. All these issues are beyond the scope of the present paper.

As stressed in \cite{MC}, the laws of physics can be the usual ones even in fractal sets, provided one knows the explicit form of $f$. The choice of $f$ may be, though, restricted by some additional laws, such as the thermodynamic formalism we have just outlined. For the moment such additional laws are unknown.

\section{Irregularities of $C$ violate parity invariance at large resolutions}

The readers may have noticed that for $\ominus x\neq -x$, i.e. $-f(x)\neq f(-x)$, one implicitly violates parity invariance, a property that leads to a reasonable estimate of $\ell<10^{-18}$m, which is the electroweak range. Plots such as those from Figs.~1, 2, and 4 show that an antisymmetric $f$ implies an unphysical-looking symmetry around $x^1=0$ of space-time fractals. According to the Copernican principle no preferred $x^1$ should be {\it a priori\/} assumed. This can be achieved either by translation invariance of space, which is excluded if a fractal structure is present, or  by a complete irregularity of $f$. This is the main reason why the notion of multi-resolution Cantor line is introduced.  Fig.~8 and Fig.~9 show an example of $f$ constructed by means of a slightly less trivial $p_k^{(n,j)}$. Here we have chosen
\be
p_0^{(n,j)} &=& p_2^{(n,j)}=\frac{1}{2}\left(1-\frac{1}{3(|j+1|+1)}\right),\\
p_1^{(n,j)} &=& \frac{1}{3(|j+1|+1)}.
\ee
Independence of $n$ makes the $n\to\infty$ limit trivial but the effective dimension is $j$ dependent. Solving $\left(p_0^{(n,j)}\right)^{D^{(n,j)}}+\left(p_2^{(n,j)}\right)^{D^{(n,j)}}=1$ for $D^{(n,j)}$ we find that the minimal dimension $\log_32$ is for $j=-1$, and with $j\to\pm \infty$ the dimensions tend to 1. Fig.~9 shows the same plot as in Fig.~8, but from a wider perspective, illustrating the effective disappearance of irregularities at distances much larger than $\ell$.

\section{Change of physical units}

Conceptual difficulties and subtleties related to fundamental length $\ell$ are well known and have been discussed in the literature for more than a century (for a relatively recent discussion cf. \cite{DSR,JoMa,Duff}). Here we would like to make some remarks on the use of dimensional quantities in non-Diophantine arithmetic, and in fractal arithmetic in particular. The term `dimension' is here understood in relation to systems of physical units \cite{Sonin}, and not to Hausdorff dimensions or the like.
 
First of all, there exists a class of $f$s that do not lead to any difficulties with dimensional quantities, namely functions of the form $f(x)=x^q$. For such an $f$ one finds $x\odot y=xy$, $x\oplus y=(x^q+y^q)^{1/q}$, so multiplication and division remain unchanged \cite{MC}. Rules such as 1km=1000m are unaffected by the change of arithmetic. Addition is not problematic either:
\be
2\textrm{m}\oplus 3\textrm{km}
&=&
2\textrm{m}\oplus 3000\textrm{m}
=
\big((2\textrm{m})^q+(3000\textrm{m})^q\big)^{1/q}\nonumber\\
&=&
\big(2^q+3000^q\big)^{1/q}\textrm{m}=(2\oplus 3000) \textrm{m}.
\ee
Alternatively,
\be
2\textrm{m}\oplus 3\textrm{km}
&=&
\big(0.002^q+3^q\big)^{1/q}\textrm{km}=(0.002\oplus 3) \textrm{km}\nonumber\\
&=&
(0.002\oplus 3) 1000\textrm{m}=(2\oplus 3000) \textrm{m},
\ee
since $(x\oplus y)\odot z=(x\oplus y)z=(xz)\oplus (yz)$. Let us mention that quantum harmonic oscillator formulated in terms of $f(x)=x^q$ arithmetic has energy levels $E_n=\frac{\hbar\omega}{2}(2n+1)^{1/q}$ \cite{MC}.

Benioff's $f(x)=px$ rescales multiplication, $x\odot y=pxy$, but keeps addition unchanged, $x\oplus y=x+y$. Now the oscillator has energy levels $E_n=p\hbar\omega(n+1/2)$ \cite{MC}. The example shows that a change of arithmetic may have nontrivial consequences for the issue of varying fundamental constants \cite{JoMa,Duff}.

Although the above two cases have not led to difficulties with dimensional variables, this will not be so in general.
The bijection $f(x)=(x+x^3)/2$, $f:\mathbb{R}\to \mathbb{R}$, satisfies all the assumptions needed for a well defined non-Diophantine arithmetic, but applies only to dimensionless variables. An attempt of computing $f(1\textrm{km})=(\textrm{km}+\textrm{km}^3)/2$ leads, from the point of view of physics, to an ill defined expression. In fractal sets this type of difficulty will be generic.

A dimensional quantity is a pair, $(x,\ell)$ say, in standard notation denoted by $x\ell$, but $x$ and $\ell$ are not objects of the same type: $x$ is dimensionless while $\ell$ keeps track of the type of physical quantity. The fundamental unit $\ell$ plays a role of an abstract index, analogous to `Alice' and `Bob' in cryptography, or the Penrose spinor/tensor abstract indices. The change of scale by $\lambda$ is mathematically achieved by the identification $\lambda(x,\ell)=(x,\lambda\ell)=(\lambda x,\ell)$. So, dimensional quantities belong to a quotient space obtained by dividing a Cartesian product by an equivalence relation. This is in fact how in abstract algebra one defines a tensor product. We can thus say that the dimensional quantity $(x,\ell)$ is a tensor product $x\otimes \ell$. But now we deal with three different sets: The dimensionless $X=\{x\}$, the collection of all the possible fundamental lengths ${\cal L}=\{\ell\}$, and the tensor product $X\otimes {\cal L}$. In principle, in each of these sets we can define different arithmetic operations, provided they are mutually consistent. So let $\odot$, $\oplus$ be the operations in $X$, `$\cdot$', `$+$' be those in $\cal L$, and let 
$\tilde\odot$, $\tilde\oplus$ act in $X\otimes {\cal L}$. 
In order to identify 
\be
(\lambda \odot x)\otimes \ell=x\otimes(\lambda\cdot \ell)=\lambda\tilde\odot (x\otimes \ell)
\ee
we have to use only such $\lambda$s that $\lambda \odot x$ and $\lambda\cdot \ell$ simultaneously make sense. 
For example, in the Cantor line $C$ introduced in \cite{MC} one finds $1/3\in C$ and $2/3\notin C$. The change of units $\ell\to\ell/3$ is then meaningful, but $\ell\to 2\ell/3$ is not. 

An interesting exercise is to solve the energy eigenvalue problem for the quantum harmonic oscillator with the non-Diophantine arithmetic defined by $f(x)=(x+x^3)/2$, and then link dimensionless parameters with observable quantities. This could be done along the lines described in \cite{MC}, but would lead us too far astray from the main topic of the present paper. A detailed discussion will be presented elsewhere.

\section{Further physical implications of relativity of arithmetic}

The principle of relativity of arithmetic \cite{MC} states that the usual laws of physics do not tell us which $f$ to choose in order to define `the physical' arithmetic operations. Perhaps Ockham razor is in order, and $f(x)=x$ should be selected for reasons of simplicity, or maybe some new physical laws are needed. Alternatively, what we perceive as physical quantities may not be the elements of $X$, equipped with $\odot$ and $\oplus$, but rather their images $f(X)$ where `$+$' and `$\cdot$' apply (cf. Fig.~5). The twin paradox in fractal space-time provides an illustration of this argument. Indeed, even if the velocity $\beta\in X$ is a fractal quantity, originating from a fractal nature of both space and time, it is the image $f(\beta)\in [-1,1]\subset \mathbb{R}$ that enters the expression $\sqrt{1-f(\beta)^2}$ describing the time delay. Perhaps this is why out of all the curves depicted in Fig.~6, what one experimentally observes is the $q=1$ case, independently of $f$. 
One can see here an analogy to the special principle of relativity, stating that there is no preferred inertial reference frame. 
This type of interpretation echoes the paradigm of multi-scale spacetimes \cite{FC2,FC3}.

Let us note, however, that the argument does not work anymore if one encounters {\it two\/} sets, $X_1$ and $X_2$, such that they cannot be simultaneously described by the same $f$. An analogy from special relativity would be the case of two inertial frames in relative motion. This is in fact what happens with the Cantor-like set $C$ embedded in $\mathbb{R}$. Even if the arithmetic of $\mathbb{R}$ is the `standard' one with $f(x)=x$, the arithmetic in the Cantor set requires a nontrivial $f$ (the Cantor-line function). Now the freedom of choosing $f$ is limited by geometric relations between the fractal set $X_1=C\subset \mathbb{R}$ and $X_2=\mathbb{R}$. In principle, having one $f$ that describes $C$ one can further modify arithmetic by applying some new bijection $g$ to both $X_1$ and $X_2$. An analogy from special relativity would be two inertial frames in relative motion, but seen from yet another frame of reference, even not of an inertial type.

Space-time fractals constructed by means of homogeneous spaces of Lie groups involving fractal arithmetic of group parameters lead to a new concept of symmetry. This is clearly seen in Fig.~1 where all the sets are {\it rotationally invariant\/}, in spite of their Cantorian appearance. Such a symmetry is `internal' in the sense that it can be identified only after having identified the implicit arithmetic of a fractal object. It is thus natural to ask if astronomical fractal-like objects, such as galaxy clusters, halos or voids, can be equipped with these `internal' symmetries. If so, what are their physical implications?

Particularly intriguing is the case of $X_1$ whose Lebesgue measure is zero. Sets of zero measure are invisible from the point of view of quantum mechanics since all wave functions that are identical {\it up to sets of zero measure\/} represent the same state. If the zero-measure set $X_1$ is equipped with an appropriate bijection $f$ (this is the case of the Cantor set), one can formulate physics (classical and quantum) within $X_1$. An example of a quantum harmonic oscillator in a Cantor line was described in \cite{MC}, with the conclusion that energy of such a system is analogous to dark energy. Obviously, all physical quantities associated with sets of measure zero will `come out of nowhere' from the point of view of standard quantum mechanics, and thus will be as `dark' as the dark energy.

\section*{Appendix: Proof of Eq.~(\ref{1=0(2)})}

As usual, the symbol `$(2)$' denotes the infinite sequence of 2s. Let us first show that
\be
r_2^{(n,j)}+p_2^{(n,j)}r_2^{(n+1,j)}+p_2^{(n,j)}p_2^{(n+1,j)}r_2^{(n+2,j)}+\dots=1.\label{2}
\ee
Fig.~7 shows that
\be
r_2^{(1,j)}+p_2^{(1,j)}r_2^{(2,j)}+p_2^{(1,j)}p_2^{(2,j)}r_2^{(3,j)}+\dots=1
\nonumber
\ee
by definition.
So
\be
1&=&r_2^{(1,j)}+p_2^{(1,j)}r_2^{(2,j)}+p_2^{(1,j)}p_2^{(2,j)}r_2^{(3,j)}+\dots\nonumber\\
&=&
1-p_2^{(1,j)}+p_2^{(1,j)}r_2^{(2,j)}+p_2^{(1,j)}p_2^{(2,j)}r_2^{(3,j)}+\dots\nonumber\\
&=&
1+p_2^{(1,j)}\Big(-1+r_2^{(2,j)}+p_2^{(2,j)}r_2^{(3,j)}+\dots\Big)\nonumber\\
&=&
1+p_2^{(1,j)}p_2^{(2,j)}\Big(-1+r_2^{(3,j)}+p_2^{(3,j)}r_2^{(4,j)}+\dots\Big)\nonumber\\
&=&
1+\Pi_{l=1}^{n-1}p_2^{(l,j)}\Big(-1+r_2^{(n,j)}+p_2^{(n,j)}r_2^{(n,j)}+\dots\Big)\nonumber
\ee
which implies (\ref{2}) for any $n$. Now consider
\be
x
&=&
x_{j.a_1\dots a_n0(2)}
\nonumber\\
&=&
j
+
s^{j}_{a_1}
+
\dots
+s^{j}_{a_1\dots a_n}
+s^{j}_{a_1\dots a_n0(2)}
\nonumber\\
&=&
j
+
s^{j}_{a_1}
+
\dots
+s^{j}_{a_1\dots a_n}
\nonumber\\
&\pp=&
+
 p_{a_1}^{(1,j)}\dots p_{a_{n}}^{(n,j)}p_{0}^{(n+1,j)}r_{2}^{(n+2,j)}
\nonumber\\
&\pp=&
+
 p_{a_1}^{(1,j)}\dots p_{a_{n}}^{(n,j)}p_{0}^{(n+1,j)}p_{2}^{(n+2,j)}r_{2}^{(n+3,j)}
+\dots
\nonumber\\
&=&
j
+
s^{j}_{a_1}
+
\dots
+s^{j}_{a_1\dots a_n}
\nonumber\\
&\pp=&
+
 p_{a_1}^{(1,j)}\dots p_{a_{n}}^{(n,j)}p_{0}^{(n+1,j)}
\nonumber\\
&\pp=&
\times\Big(
r_{2}^{(n+2,j)}
+
p_{2}^{(n+2,j)}r_{2}^{(n+3,j)}
+\dots
\Big)\nonumber
\ee
Employing (\ref{2}) and $p_{0}^{(n+1,j)}=r_{1}^{(n+1,j)}$ we get
\be
x
&=&
x_{j.a_1\dots a_n0(2)}
\nonumber\\
&=&
j
+
s^{j}_{a_1}
+
\dots
+s^{j}_{a_1\dots a_n}
\nonumber\\
&\pp=&
+
 p_{a_1}^{(1,j)}\dots p_{a_{n}}^{(n,j)}r_{1}^{(n+1,j)}
\nonumber\\
&=&
j
+
s^{j}_{a_1}
+
\dots
+s^{j}_{a_1\dots a_n}
+s^{j}_{a_1\dots a_n1}
\nonumber\\
&=&
x_{j.a_1\dots a_n1}
\nonumber
\ee
which ends the proof.

\end{document}